\documentclass[11pt,twoside]{article} 

\usepackage{fullpage}

\RequirePackage{amsthm,amsmath,amsfonts,amssymb}
\RequirePackage{natbib}

\usepackage{graphics} 
\usepackage{tikz}
\usetikzlibrary{quantikz}
\usetikzlibrary{arrows}
\usetikzlibrary{shapes,fadings,snakes}
\usetikzlibrary{decorations.pathmorphing,patterns}
\usetikzlibrary{calc}
\usepackage{hyperref}
\usepackage{cleveref}
\usepackage{epsfig}
\usepackage{graphicx}
\usepackage{subcaption}
\usepackage{xcolor}
\usepackage{physics}
\usepackage[most]{tcolorbox}
\usepackage{bbm}
\usepackage{overpic}
\usepackage{diagbox}
\usepackage{lineno}

\newtheorem{proposition}{Proposition}
\newtheorem{assumption}{Assumption}

\theoremstyle{remark}
\newtheorem{remark}{Remark}

\newcommand\independent{\protect\mathpalette{\protect\independenT}{\perp}}
\def\independenT#1#2{\mathrel{\rlap{$#1#2$}\mkern2mu{#1#2}}}

\newcounter{mybox}
\refstepcounter{mybox}
\def\pr{\mathbb{P}}
\def\E{\mathbb{E}}

\def\tr{\mathrm{tr}}

\title{A quantum experiment with joint exogeneity violation}

\author{Yuhao Wang$^{1,2, 3}$\thanks{correspondence should be addressed to YW: \href{mailto:yuhaow@tsinghua.edu.cn}{yuhaow@tsinghua.edu.cn}}\; and Xingjian Zhang$^{4}$}

\date{
	$^1$Institute for Interdisciplinary Information Sciences, Tsinghua University \\
	$^2$Shanghai Qi Zhi Institute \\
	$^3$Shanghai Artificial Intelligence Laboratory \\
	$^4$QICI Quantum Information and Computation Initiative,
	School of Computing and Data Science, The University of Hong Kong\\
	\vspace{1em}
	Prelimiary Draft\thanks{This is a preliminary draft in accordance to YW’s presentation at the 2025 Pacific Causal Inference Conference and 2025 Chinese Causal Inference Conference. 
    This draft is being circulated to collect further feedback. A formal version will be publicly available in due course.
    } \\
	\vspace{1em}
	\today
}

\begin{document}

	\maketitle

\begin{abstract}
In randomized experiments, the assumption of potential outcomes is usually accompanied by the \emph{joint exogeneity} assumption. Although joint exogeneity has faced criticism as a counterfactual assumption since its proposal, no evidence has yet demonstrated its violation in randomized experiments. In this paper, we reveal such a violation in a quantum experiment, thereby falsifying this assumption, at least in regimes where classical physics cannot provide a complete description. We further discuss its implications for potential outcome modelling, from both practial and philosophical perspectives.
\end{abstract}


\section{Introduction}

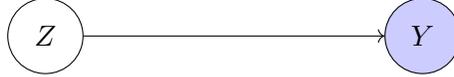
\begin{figure}[t!]
	\centering 
	\begin{tikzpicture}
		\node[draw, circle, text centered, minimum size = 1cm] (Z) {$Z$};
		\node[draw, circle, text centered, right=4cm of Z, minimum size = 1cm, fill=blue!20] (Y) {$Y$};
		
		\draw[->] (Z) -- (Y);
	\end{tikzpicture}
	\caption{A directed acyclic graph representing a simple experiment. Here $Z \in \{0, 1\}$ represents the treatment assignment, and $Y$ represents the outcome.}
	\label{fig:experiment}
\end{figure}

Inferring causality from data is a central topic in scientific research. 
To successfully model causal effects, a common approach is to use \emph{potential outcomes}~\citep{imbens2015causal,rubin2005causal}. To illustrate the concept of potential outcomes, consider a simple experiment where the treatment assignment $Z \in \{0,1\}$ is assigned uniformly at random, and we observe an outcome random variable $Y$ (see e.g.~\Cref{fig:experiment}). We can associate $Y$ with two potential outcomes, $Y(1)$ and $Y(0)$, representing the outcomes that would have been observed if the individual had been assigned to treatment group $1$ or $0$, respectively.
Based on this definition, the relationship between potential outcomes and the observed outcomes satisfy that $Y = Z Y(1) + (1 - Z) Y(0)$. The potential outcome framework lays as the foundations of causal modelling. This is also called \emph{Rubin Causal Model}, or \emph{Neyman-Rubin Causal Model}~\citep{imbens2015causal}.

Armed with the potential outcomes framework, the next step is to identify causal effects from the observed data distribution. Successful causal effect identification requires key assumptions. Among these, the exogeneity assumption (also called unconfoundedness) forms the foundation of causal effect identification~\citep{imbens2015causal}. In the context of a randomized experiment where treatment is assigned uniformly at random, the existing mathematical formulation of exogeneity is mainly split into two types: marginal exogeneity and joint exogeneity. 

Marginal exogeneity assumes that the treatment assignment should be independent of each potential outcome. For example, returning to the simplest experiment in \Cref{fig:experiment} for illustration, marginal exogeneity means that $Y(z) \independent Z$ for all $z \in \{0,1\}$. Joint exogeneity, on the other hand, requires not only that the treatment assignment indicator should be independent of each potential outcome, but also their joint vector: that $(Y(1), Y(0)) \independent Z$.
Joint exogeneity is sometimes criticized as more restrictive than marginal exogeneity, as it constrains the cross-world distribution $\mathbb{P}(Z, Y(1), Y(0))$, which is a fundamentally unobservable quantity even in hypothetical experiments. 
Nevertheless, if one holds the view that potential outcomes are deterministic or exist as latent variables generated \emph{before} the experiment, and conducting a randomized experiment merely reveals one pre-existing outcome, joint exogeneity remains a reasonable assumption.
Since in this case, $Z$, whose randomness stems solely from the experimenter, should naturally be independent of $(Y(1), Y(0))$.
\citet{dawid2000causal} terms this a ``fatalism'' philosophical stance, considering it controversial because it is ``metaphysical'' and, in his own words:
\begin{quote}
 ``counter to the philosophy underlying statistical modeling and inference in almost every other setting''.
\end{quote}



Though joint exogeneity is an controversial assumption since it originates from an attidude of fatalism, at least to our knowledge, until the moment this manuscript becomes publically available, there is still no known randomized experiment that violates joint exogeneity.
In this paper, we provide a quantum experiment, where assuming joint exogeneity can result in a contradication, thereby falsifying this assumption. We believe this finding has many implications. In particular, it challenges the fatalism view that potential outcomes pre-exist before treatment is assigned. Instead, a more libertarianism view that the potential outcomes are generated concurrently or immediately \emph{after} the treatment assignment, whose joint distribution may depend on experimenter's choice of $Z$, is more convincing. Moreover, this work uses a real and implementable quantum system to rigorously justify the suggestion from \citet{gill2014statistics,robins2015proof} regarding the choice between realism and locality in potential outcome modelling.

The rest of this paper is organized as follows. In \Cref{sec:jevio}, we present the example where the joint exogeneity is violated. In \Cref{sec:theory}, we analyze the main theoretical conclusions derived in \Cref{sec:jevio}.
In \Cref{sec:pomodelling}, we show the implications to our understanding of the potential outcome framework. 
We end this paper with a conclusion in~\Cref{sec:conclusion}.

\section{An example for joint exogeneity violation}\label{sec:jevio}

%

\begin{figure}[t!]
	\centering 
    \begin{subfigure}{0.4\textwidth}
    	\centering
    \resizebox{!}{2.3cm}{%
	\begin{tikzpicture}
		\node[draw, circle, text centered, minimum size = 1cm, fill=blue!20] (X) {$X$};
		\node[draw, circle, text centered, right=4cm of X, minimum size = 1cm, fill=blue!20] (Y) {$Y$};
		\node[draw, regular polygon, regular polygon sides=3, text centered, above right=1.5cm and 2cm of X, fill=brown!20] (L) {$\Lambda$};
		\node[draw, circle, text centered, right=4cm of X, minimum size = 1cm, above left=1.5cm and 2cm of X] (Z) {$Z$};
		
		\draw[->] (X) -- (Y);
		\draw[->] (L) -- (Y);
		\draw[->] (L) -- (X);
		\draw[->] (Z) -- (X);
		\draw[->, red] (Z) -- (Y);
	\end{tikzpicture}
    }
    \caption{Quantum network representation}
    \label{fig:instrument}
    \end{subfigure}
    \hfill
    \begin{subfigure}{0.55\textwidth}
    	\centering
        \resizebox{!}{2.3cm}{%
	\begin{tikzpicture}
		\node[draw, circle, text centered, minimum size = 1cm, fill=blue!20] (X) {$X$};
		\node[draw, circle, text centered, right=4cm of X, minimum size = 1cm, fill=blue!20] (Y) {$Y$};
		\node[draw, regular polygon, regular polygon sides=3, text centered, above right=1.5cm and 2cm of X, fill=brown!20] (L) {$\Lambda$};
		\node[draw, circle, text centered, right=4cm of X, minimum size = 1cm, above left=1.5cm and 2cm of X] (Z) {$Z$};
        \node[draw, rectangle, text centered, minimum size = 1cm, above right =1.5cm and 2cm of Y] (ZZ) {$(X = x,Y = z)$};
		
		\draw[->] (X) -- node[pos=0.4]{\huge $\times$} (Y);
		\draw[->] (L) -- (Y);
		\draw[->] (L) -- (X);
		\draw[->] (Z) -- (X);
		\draw[->, red] (Z) -- node[pos=0.4]{\huge $\times$} (Y);
        \draw[->] (ZZ) -- (Y);
	\end{tikzpicture}
    }
    \caption{Representation of the experiment in~\Cref{sec:jevio}}
     \label{fig:intervention}
    \end{subfigure}
	\caption{(a) A directed acyclic graph representing the quantum communication network considered in \Cref{sec:jevio}. The blue nodes represent the endogenous nodes, the transparent node represents the exogeneous nodes; the triangle represents the quantum state. The red line from $Z$ to $Y$ means that the measurement at node $Y$ will be performed after receiving the signal from $Z$. However, the choice of the measurement operator will not depend on the realization of $Z$. (b) Representation of the hypothetical experiment described in~\Cref{sec:jevio}. Here, we assume that after random variables $X, Z$ are generated, instead of sending the observed value to $Y$, it instead resets them fixed values $x, z \in \{0, 1\}$ (denoted by the rectangle on the right), and sends the new values node $Y$.}
\end{figure}
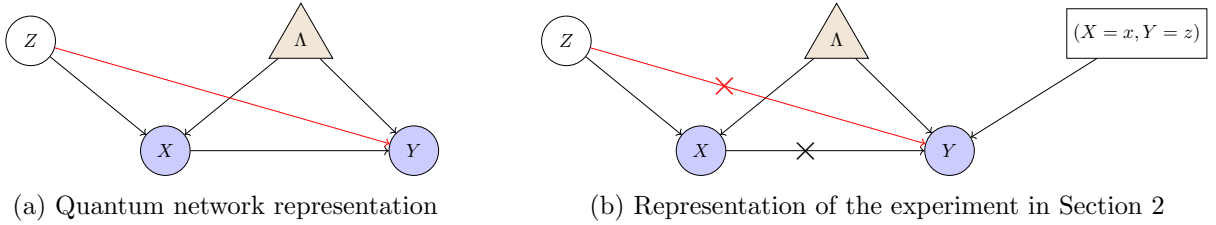

Consider a quantum communication network according to~\Cref{fig:instrument}, where $Z, X, Y$ are observed binary classical random variables. $Z$ is a random variable whose distribution is fully determined by the experimenter. $\Lambda$ represents a quantum system, and $X$ and $Y$ represent two measurements of the quantum system $\Lambda$. Specifically, at node $X$, the experimenter first receives the signal from $Z$, and performs a measurement of the quantum system based on the realized value of $Z$. At node $Y$, the experimenter receives the signals from $X$ and $Z$. However, when performing the measurement after receiving signals from $Z$ and $X$, the measurement operator choice at $Y$ depends \emph{only} on the output from $X$, not on $Z$. As a concrete physical realization, for example,~\Cref{fig:instrument} can represent a photonic experiment, where the experimenter determines $Z$, and $\Lambda$ is a light source sending two photons with entangled polarizations to nodes $X$ and $Y$. The measurement of $X$ is determined by the realization of $Z$, which can be physically implemented by adjusting the angle of a polarizer at node $X$ based on the output from $Z$.
Similarly, for node $Y$, the experimenter first receives signals from nodes $X$ and $Z$, and then passes the photon received from the light source through a polarizer, whose angle depends only on the output from $X$, not $Z$, and then measures the photon after that.

We let $\rho$ denote the density operator of the quantum state $\Lambda$, and let $M_x^z$ and $N_y^x$ denote the measurement operators of the measurements at nodes $X$ and $Y$, respectively, then the conditional distribution of $X, Y$ given $Z$ can be expressed as (see also~\citet{chaves2018quantum}):
\begin{equation}\label{eq:QProb}
    \pr(X = x, Y = y \mid Z = z) = \tr[(M_x^z \otimes N_y^x) \rho].
\end{equation}
Now, assuming the existence of potential outcomes $\{Y(x,z)\}_{x, z \in {0, 1}}$, the conditional distribution $\pr(Y(x, z) = y \mid X = x', Z = z')$ then represents the distribtuion of the potential response of $Y$ if $X$ were set to $x$ and $Z$ to $z$, conditional on the observed actual treatment assignment $(X = x', Z = z')$. This is sometimes also referred to as the distribution of the potential outcome under treatment arm $(x, z)$ within the subpopulation receiving $(x', z')$. 
In other words, $\pr(Y(x, z) = y \mid X = x', Z = z')$ equals the distribution of $Y$ that would be observed in a hypothetical experiment where the experimenter first generates $X$ and $Z$ according to~\Cref{fig:instrument}; then under the event $(X = x', Z = z')$, intead of intervening $Y$ according to $(x', z')$, the experimenter instead resets $(X, Z)$ to predetermined values $(x, z)$, and performs the intervention afterwards (see the following paragraph for more details). 
This definition forms the basis of key causal inference concepts, including the average treatment effect on the treated (ATT) and the average treatment effect on the controls (ATC). 

We now investigate the properties of $\pr(Y(x, z) = y \mid X = x', Z = z')$. As discussed in the previous paragraph, the joint distribution of $Y(x, z), X, Z$ is equivalent to the interventional distribution of $Y, X$ and $Z$ when they are generated according to the following thought experiment. First, the experimenter randomly draws a $Z$ and measures $X$ following the same protocol as the one in~\Cref{fig:instrument}, and records their values. Second, when measuring at node $Y$, instead of sending the actual $X$ and $Z$ to node $Y$, the experimenter resets them to predetermined values $x, z \in \{0, 1\}$, and then sends the new values to the node, and then measures the quantum system by adjusting the measurement operator based on the these new values.  
See~\Cref{fig:intervention} for a graphical illustration. Write $\mathbb{Q}_{xz}$ as the distribution of random variables $X, Y, Z$ under this interventional experiment, then by definition, $\mathbb{Q}_{xz}(Y = y \mid X = x', Z = z') = \pr(Y(x,z) = y \mid X = x', Z = z')$. Moreover, according to standard results in quantum theory,

\begin{equation}\label{eq:qxz}
    \mathbb{Q}_{xz}(X = x',Y = y \mid Z = z') = \tr[(M_{x'}^{z'} \otimes N_y^x) \rho],
\end{equation}
where the right hand side does not depend on $z$.
Putting together, we obtain that for any $x, y, x', z' \in \{0,1\}$,
\begin{equation}\label{eq:ser}
	\pr(Y(x,1) = y \mid X = x', Z = z') = \pr(Y(x,0) = y \mid X = x', Z = z').
\end{equation}
We call this as ``\emph{stratified} exclusion restriction''.

Now we invoke the additional joint exogeneity assumption, which is the only assumption in addition to the existence of potential outcomes:
\begin{assumption}[joint exogeneity]\label{as:jeiv}
    $Z \independent (Y(0, 0), Y(0, 1), Y(1, 0), Y(1, 1))$.
\end{assumption}
In the following proposition, we show that the involvement of~\Cref{as:jeiv} results in the famous Balke-Pearl bound~\citep{balke1997bounds}.
\begin{proposition}\label{prop:bpb}
	Consider the quantum experiment in~\Cref{fig:instrument}. Suppose the potential outcomes $\{Y(x, z)\}_{x, z \in \{0,1\}}$ exist, and in addition~\Cref{as:jeiv}, then we have that $\E(Y(1,z) - Y(0,z))$ is within the Balke-Pearl bound.
\end{proposition}


The proof of~\Cref{prop:bpb} relies on showing that the average causal effect is identifiable up to Balke-Pearl bound by assuming joint exogeneity and the so-called stratified exclusion restriction assumption~\eqref{eq:ser}, with the latter directly implied by the existence of potential outcome in the quantum system described in~\Cref{fig:instrument}. We would like to remark that these are not the only combinations of assumptions that makes the average causal effect identifiable up to the Balke-Pearl bound. There are also other assumptions that delivers the same degree of identification. For more information, we refer the readers to~\citet{swanson2018partial} and~\citet[Chapter~5]{guo2021}.


However, as shown in~\cite{chaves2018quantum}, there may exist some $M_z^x, N_x^y$ and $\rho$ such that the corresponding true causal effect is outside the range of Balke-Pearl bound, which raises a contradiction. 
Since there are two assumptions underlie \cite{chaves2018quantum}: first, the existence of potential outcomes denoted as $\{Y(x,z)\}_{x, z \in \{0,1\}}$; and second, joint exogeneity. The contradiction then implies that at least one of the two assumptions are invalid. Either scenario poses a critical issue for invoking~\Cref{as:jeiv}. In other words, this contradiction falsifies \Cref{as:jeiv} in this experiment. In~\Cref{sec:pomodelling}, we will provide more discussions regarding this violation. We end this section with the following remark.

\begin{remark}
Following standard derivations in quantum theory, we can have from~\eqref{eq:qxz} that $\mathbb{Q}_{xz}(Y = y \mid Z = 1) = \mathbb{Q}_{xz}(Y = y \mid Z = 0)$, namely that $\pr(Y(x,z) = y \mid Z = 1) = \pr(Y(x,z) = y \mid Z = 0)$, thereby justifying the marginal exogeneity assumption. 
\end{remark}


\section{Theoretical analysis}\label{sec:theory}
In this section, we show how the results in~\Cref{prop:bpb} is derived. Above all, assuming the existence $Y(x,z)$, we can relate them with the observables $X,Y,Z$ via the following equation,
\begin{equation}\label{eq:Def}
  Y=XZY(1,1)+(1-X)ZY(0,1)+X(1-Z)Y(0,1)+(1-X)(1-Z)Y(0,0).
\end{equation}
Here, all the variables take binary values from $\{0,1\}$.
Therefore, for $x,y,z \in \{0,1\}$,
\begin{equation}\label{eq:obs}
  \pr(X=x,Y=y|Z=z)=\pr(X=x,Y(x,z)=y|Z=z),
\end{equation}
where the left-hand side (LHS) denotes the observed distribution of the quantum network. For simplicity, we denote $p_{XY|Z}(x,y|z)\equiv\pr(X=x,Y=y|Z=z)$. 
We are interested in the average causal effect (ACE) from $X$ to $Y$, since we assume throughout that the potential outcomes exist, we can define it via
\begin{equation}\label{eq:ACE}
  \mathrm{ACE}_{X\rightarrow Y}\equiv \pr(Y(1,0) = 1)-\pr(Y(0,0) = 1) \equiv \pr(Y(1,0) = 1 \mid Z = 0)-\pr(Y(0,0) = 1 \mid Z = 0),
\end{equation}
where the last inequality is directly from~\Cref{as:jeiv}. Our goal is to derive bounds on ACE, given the constraint that the observed distributions follow the prespecified $p_{XY|Z}(x,y|z)$. To derive these bounds, we will express the target function and constraints via the conditional probabilities, 
\[
\pr(X = x,Y(0,0) = y_{00}, Y(0,1) = y_{01}, Y(1,0)  = y_{10}, Y(1,1) = y_{11}| Z = z) \forall x, y_{00}, y_{01}, y_{10}, y_{11}, z \in \{0, 1\},
\]
which can be seen as a $2^6=64$-dimensional vector, and then transform the bound derivations as solving linear programs, just like~\citet{balke1997bounds}. For later convenience, we abbreviate the notations as
\[
q_{x,y_{00},y_{01},y_{10},y_{11}|z}\equiv\pr(X=x,Y(0,0)=y_{00},Y(0,1)=y_{01},Y(1,0)=y_{10},Y(1,1)=y_{11}|Z=z).
\]

The target function for the linear program is provided in~\eqref{eq:ACE}, we now discuss the constraints.
First, the definition of conditional probabilities lead to the following constraints: (1) non-negativity, that $\forall x,y_{00},y_{01},y_{10},y_{11},z\in\{0,1\}$,
\begin{equation}\label{eq:prob}
  \pr(X=x,Y(0,0)=y_{00},Y(0,1)=y_{01},Y(1,0)=y_{10},Y(1,1)=y_{11}|Z=z)\geq0,
\end{equation}
and (2) normalization, that $\forall z\in\{0,1\}$,
\begin{equation}
  \sum_{x,y_{00},y_{01},y_{10},y_{11}} \pr(X=x,Y(0,0)=y_{00},Y(0,1)=y_{01},Y(1,0)=y_{10},Y(1,1)=y_{11}|Z=z)=1.
\end{equation}
There are hence 64 inequality constraints and 2 equality constraints arising from the definition of the conditional probability. 
Second, under the assumption of joint exogeneity (Assumption~\ref{as:jeiv}), we can derive the constraints
\begin{equation}
\begin{aligned}
    & \pr(Y(0,0)=y_{00}, Y(0,1)=y_{01}, Y(1,0)=y_{10}, Y(1,1)=y_{11} | Z=1) \\
    & \quad \equiv \pr(Y(0,0)=y_{00}, Y(0,1)=y_{01}, Y(1,0)=y_{10}, Y(1,1)=y_{11} | Z=0)\ \forall y_{00}, y_{01}, y_{10}, y_{11} \in \{0, 1\}.
\end{aligned}    
\end{equation}
Finally, assuming existence of potential outcomes directly implies stratified exclusion restriction, yielding
\begin{equation}\label{eq:serconst}
    \pr(Y(x,1)=y, X = x' \mid Z = z')=\pr(Y(x,0)=y, X = x' \mid Z = z'),\forall x, y, x', z'\in\{0,1\}.
\end{equation}
By expressing the target function~\eqref{eq:ACE} and the constraints \eqref{eq:obs}, \eqref{eq:prob}--\eqref{eq:serconst} using the probabilities
\[
q_{x,y_{00},y_{01},y_{10},y_{11}|z}\equiv\pr(X=x,Y(0,0)=y_{00},Y(0,1)=y_{01},Y(1,0)=y_{10},Y(1,1)=y_{11}|Z=z),
\]
we can transform the problem of calculating the upper and lower bounds of identification region through solving two linear programs. Taking the calculation of lower bound for illustration, it is equivalent to solving the following linear program:
\begin{equation}\label{eq:JELinprog}
\begin{aligned}
    \min_{\substack{q_{x,y_{00},y_{01},y_{10},y_{11}|0}, \\q_{x,y_{00},y_{01},y_{10},y_{11}|1}}} &\left\{\sum_{x,y_{10},y_{01},y_{11}}q_{x,y_{00},y_{01},1,y_{11}|0}-\sum_{x,y_{01},y_{10},y_{11}}q_{x,1,y_{01},y_{10},y_{11}|0}\right\}, \\
    \text{subject to:}& \\
    \sum_xq_{x,y_{00},y_{01},y_{10},y_{11}|0} &= \sum_xq_{x,y_{00},y_{01},y_{10},y_{11}|1}, \forall y_{00},y_{01},y_{10},y_{11}\in\{0,1\}, \text{(joint exogeneity)} \\
    \sum_{y_{01},y_{10},y_{11}}q_{x,0,y_{01},y_{10},y_{11}|z} &= \sum_{y_{00},y_{10},y_{11}}q_{x,y_{00},0,y_{10},y_{11}|z},  \forall x,z\in\{0,1\}, \\
    \sum_{y_{00},y_{01},y_{11}}q_{x,y_{00},y_{01},1,y_{11}|z} &= \sum_{y_{00},y_{01},y_{10}}q_{x,y_{00},y_{01},y_{10},1|z},  \forall x,z \in \{0,1\}\text{(stratified exclusion restriction)}\\
    \sum_{x,y_{00},y_{01},y_{10},y_{11}}q_{x,y_{00},y_{01},y_{10},y_{11}|z} &=1, \forall z \in \{0,1\}, \text{(probability normalization)}\\
    \sum_{y_{01},y_{10},y_{11}}q(0,y_{00},y_{01},y_{10},y_{11}|0) &= p_{XY|Z}(0,y_{00}|0), \forall y_{00} \in \{0,1\}\\
   \sum_{y_{00},y_{01},y_{11}}q(1,y_{00},y_{01},y_{10},y_{11}|0) &= p_{XY|Z}(1,y_{10}|0), \forall y_{10} \in \{0,1\}\\
   \sum_{y_{00},y_{10},y_{11}}q(0,y_{00},y_{01},y_{10},y_{11}|1) &= p_{XY|Z}(0,y_{01}|1), \forall y_{01} \in \{0,1\}\\
   \sum_{y_{00},y_{01},y_{10}}q(1,y_{00},y_{01},y_{10},y_{11}|1) &= p_{XY|Z}(1,y_{11}|1), y_{11} \in \{0,1\}\text{(Observations)} \\
   q_{x,y_{00},y_{01},y_{10}|z}&\geq 0, \forall x,y_{00},y_{01},y_{10},z \in \{0,1\}. \text{(Probability non-negativity)}
\end{aligned}
\end{equation}
This linear programming can be handled analytically, which yields the same bounds as the so-called Balke-Pearl bound, thereby proving \Cref{prop:bpb}.

In \cite{chaves2018quantum}, the authors show that quantum theory can violate the predictions by Balke-Pearl bounds. The standard formulation of Balke-Pearl bounds assume both joint exogeneity and individual exclusion restriction. Here, our results show that even when individual exclusion restriction is relaxed to the so-called ``stratified exclusion restriction, the ACE estimation is still false in the quantum theory in general. Specifically, consider that the joint probability of $p_{XY|Z}(x,y|z)$ comes from the measurement on the Bell state, $\rho=\ketbra{\Phi^-}$ with $\ket{\Phi^-}=(\ket{00}-\ket{11})/\sqrt{2}$, with the measurements being
\begin{equation}
\begin{aligned}
    &Z=0: M^{z=0}_{x=0}-M^{z=0}_{x=1}=\sigma_Z\equiv A_0, \\
    &Z=1: M^{z=1}_{x=0}-M^{z=1}_{x=1}=\sigma_X\equiv A_1, \\
    &X=0: N^{x=0}_{y=0}-N^{x=0}_{y=1}=(\sigma_Z+\sigma_X)/\sqrt{2}\equiv B_0, \\
    &X=1: N^{x=1}_{y=0}-N^{x=1}_{y=1}=(\sigma_Z-\sigma_X)/\sqrt{2}\equiv B_1, \\
\end{aligned}
\end{equation}
where $\sigma_Z$ and $\sigma_X$ are the Pauli observables:
\[
    \sigma_Z=\begin{pmatrix}
        1 & 0 \\
        0 & -1
    \end{pmatrix},
    \sigma_X=\begin{pmatrix}
        0 & 1 \\
        1 & 0
    \end{pmatrix},
\]
and the measurement outcomes of $0$ and $1$ correspond to the $+1$ and $-1$ eigenspaces of the observables $A_0,A_1,B_0,B_1$, respectively. On the computational basis, the Bell state is represented as the vector
\[
\ket{\Phi^-}=\frac{1}{\sqrt{2}}\begin{pmatrix}
    1 & 0 & 0 & -1
\end{pmatrix}^{\mathrm{T}}.
\]
The measurement elements are given by
\[
M^{z=0}_{x=0}=\begin{pmatrix}
    1 & 0 \\
    0 & 0
\end{pmatrix},
M^{z=0}_{x=1}=\begin{pmatrix}
    0 & 0 \\
    0 & 1
\end{pmatrix},
\]
\[
M^{z=1}_{x=0}=\frac{1}{2}\begin{pmatrix}
    1 & 1 \\
    1 & 1
\end{pmatrix},
M^{z=1}_{x=1}=\frac{1}{2}\begin{pmatrix}
    1 & -1 \\
    -1 & 1
\end{pmatrix},
\]
\[
N^{x=0}_{y=0}=\frac{1}{4-2\sqrt{2}}
\begin{pmatrix}
    1 & \sqrt{2}-1 \\
    \sqrt{2}-1 & 3-2\sqrt{2}
\end{pmatrix},
N^{x=0}_{y=1}=\frac{1}{4+2\sqrt{2}}
\begin{pmatrix}
    1 & -\sqrt{2}-1 \\
    -\sqrt{2}-1 & 3+2\sqrt{2}
\end{pmatrix},
\]
\[
N^{x=1}_{y=0}=\frac{1}{4-2\sqrt{2}}
\begin{pmatrix}
    1 & 1-\sqrt{2} \\
    1-\sqrt{2} & 3-2\sqrt{2}
\end{pmatrix},
N^{x=1}_{y=1}=\frac{1}{4+2\sqrt{2}}
\begin{pmatrix}
    1 & \sqrt{2}+1 \\
    \sqrt{2}+1 & 3+2\sqrt{2}
\end{pmatrix}.
\]
We can calculate the probabilities of $p_{XY|Z}(x,y|z)$ according to Eq.~\eqref{eq:QProb}. The probabilities are given by:
\begin{equation}
\begin{aligned}
    p_{XY|Z}(00|0)=0.4268, p_{XY|Z}(01|0)=0.0732, p_{XY|Z}(10|0)=0.0732, p_{XY|Z}(11|0)=0.4268, \\
    p_{XY|Z}(00|1)=0.0732, p_{XY|Z}(01|1)=0.4268, p_{XY|Z}(10|1)=0.0732,p_{XY|Z}(11|1)=0.4268.
\end{aligned}
\end{equation}
Taking the probabilities into the linear programming in Eq.~\eqref{eq:JELinprog}, we can derive that the estimated lower bound of ACE is $\approx0.1339$. However, as analyzed by \cite{chaves2018quantum}, the true value of ACE in this case is $0$. In other words, the quantum measurements violate the counterfactual assumptions of joint exogeneity $+$ stratified exclusion restriction.

\section{Implications for potential outcome modelling}\label{sec:pomodelling}

In this section, we discuss some practical implications of the conclusions in~\Cref{sec:jevio}. First, to our knowledge, this work provides the first example where the assumption of \emph{only} joint exogeneity can lead to a contradiction. We emphasize that this does not invalidate prior results relying on joint exogeneity, as those are primarily rooted in classical physics. Instead, we mainly aim to argue that in situations where quantum effects are non-negligible, one should be careful about invoking joint exogeneity. This finding is not only scientifically novel but also potentially relevant to real-world applications, given the rapid advancements in quantum communication and computation~\citet{lu2022micius}. Consequently, researchers, particularly in future social science field experiments, may need to critically evaluate the validity of joint exogeneity assumption, when quantum devices become common in everyday life.

In the following two subsections, we further discuss the philosophical implications from this work, extending the discussions from~\citet{dawid2000causal} and~\citet{robins2015proof,gill2014statistics}, respectively.

\subsection{Fatalism in potential outcome modelling}

Our findings provide new evidence that challenges the implicit attitude towards fatalism often held by many causal inference researchers. Although rarely stated explicitly, for a long time, and continuing to the present, many researchers (including the authors of this work) have operated under the assumption that within the potential outcomes framework, potential outcomes must exist before treatment assignment. In this view, treatment selection simply reveals one of these pre-existing potential outcomes. This perspective is notably reflected in Professor Phillip Dawid's influential critique, where he states in~\citet[Section~7]{dawid2000causal} that
\begin{quote}
	``Many counterfactual analyses are based, explicitly or implicitly, on an attitude that I term fatalism. This considers the various potential responses $Y_i(u)$, when treatment $i$ is applied to unit $u$, as predetermined attributes of unit $u$, waiting only to be uncovered by suitable experimentation.''
\end{quote}
Since random variable $Z$'s randomness stems solely from the experimenter's actions, if we consider $Y(x,z)$ either as deterministic or as random variables associated with events occurring \emph{prior} to the realization of treatment assignment $Z$, then the joint exogeneity assumption appears natural. Conversely, the violation of this assumption suggests that, at least in quantum contexts, we should not view potential outcomes as pre-existing random variables. This finding provides new empirical support for Professor Phillip Dawid's critique.

Our finding suggests that we should instead view potential outcomes as random variables that are generated \emph{concurrently with} or \emph{after} the treatment assignment, whose joint distribution \emph{depends} on experimenter's choice $Z$. Nevertheless, this does not invalidate previous results based on such a deterministic view, as they were all derived by implicitly assuming that the physical roles are classical. Our key argument is that in quantum systems where quantum phenomena become non-negligible, researchers should reconsider the predeterministic nature of the potential outcome framework.

\subsection{Locality and realism in potential outcome modelling}
\label{sec:comparison}


Notably, our paper is not the first to discuss violations of counterfactual assumptions in quantum systems. An earlier contribution is~\citet{robins2015proof}, focusing on counterfactual violations in the so-called Bell experiment. Through their analysis, \citet{robins2015proof} establishes a key philosophical insight: potential outcome models cannot simultaneously satisfy both realism 
and locality. 

To elaborate, recall our example in \Cref{fig:instrument}: realism implies that the potential outcomes should be real, pre-existing properties of units, rather than being constructed after the treatment is applied. Therefore, their joint distribution should not be affected by experimenter's free choice of $Z$ (joint exogeneity). Meanwhile, locality implies $Y(x, 1) = Y(x, 0)$ almost surely for all $x$ (i.e., individual exclusion restriction). The philosophical insight from \citet{robins2015proof} means that both joint exogeneity and individual exclusion restriction cannot hold simultaneously in quantum systems like that of~\Cref{fig:instrument}.
Consequently, \citet{robins2015proof,gill2014statistics} further recommends abandoning realism while preserving locality, as this is consistent with Copenhagen standpoints~\citep{robins2015proof}, and is in line with Occam's principle~\citep{gill2014statistics}. 

In contrast, our result demonstrates that: realism cannot be preserved within potential outcome modelling, regardless of whether locality holds. This follows because by merely assuming the existence of potential outcomes, we can derive that the joint distribution of these potential outcomes can be affected by experimenter's choice ($Z$), thereby violating the realism principle, even without invoking locality. In other words, our work provides a concrete example that rigorously justifies the recommendations from~\citet{robins2015proof,gill2014statistics} to abandon realism, resolving the realism–locality controversy within the potential outcomes framework.

\section{Conclusion}
\label{sec:conclusion}

We present a thought experiment that illustrates how the joint exogeneity assumption can be violated in quantum systems, and we examine the implications for causal modelling from both practical and philosophical standpoints. These violations indicate the need for a causal framework consistent with both classical and quantum theories -- a research direction we plan to explore in future work. \citet{zhang2024physics} offers an interesting first step; however, a complete, mature model remains an open challenge.

\bibliographystyle{plainnat}
\bibliography{bibCausal}

\end{document}